# Services décentralisés, robustes et efficaces pour une gestion autonome et temp-réel de situations d'urgences urbaines

**Frédéric Le Mouël[a*], Carlos J. Barrios Hernández[b], Oscar Carrillo[a*], Gabriel Pedraza[b]**

[a] Université de Lyon, INSA de Lyon, Lyon, France

[b] Université Industrielle de Santander, Bucaramanga, Colombia

*E-mail: prénom.nom@insa-lyon.fr

## Resumé

La mondialisation des échanges et l'organisation du travail provoquent actuellement un flux migratoire important vers les villes. Cette croissance des villes nécessite de nouvelles planifications urbaines dans lesquelles le numérique prend une place de plus en plus prépondérante - la captation des données permettant de comprendre et décider face aux changements. Ces environnements numériques sont toutefois malmenés en cas de crises (catastrophes naturelles, terrorisme, accidents, etc). Basés sur l'expertise des laboratoires CITI de l'INSA de Lyon et SC3 de l'Université Industrielle de Santander, nous nous proposons de créer le projet ALERT - *Autonomous Liable Emergency service in Real-Time*. Avec des services décentralisés, fiables et efficaces, qui soient au plus proche des citoyens, les prises de décisions pourront s'effectuer en temps réel, localement, de manière pertinente sans risque de déconnexion avec une autorité centrale. Ces collectes d'informations et prises de décision mettront en jeu la population avec des approches participatives et sociales.

**Mots clés :** Services décentralisés, Ville intelligente, Urgences.

## Introduction

Un des services citoyens présentant des enjeux importants dans les villes intelligentes est celui permettant de gérer les situations d'urgences. Nous présentons le Projet ALERT - *Autonomous Liable Emergency service in Real-Time* - un service citoyen pour les situations d'urgence.

Une urgence fait référence à une situation où des décisions doivent être prises dans de brefs délais, les conséquences de ces décisions pouvant être vitales (tremblement de terre, un attentat, etc.)

Les plateformes numériques sont essentielles dans ce genre de situation pour permettre de collecter des données de manière à caractériser la situation d'urgence et à réagir rapidement en prenant une décision. Ces situations d'urgence pouvant elles-mêmes mettre à mal les outils numériques.

Fort de l'expertise des deux laboratoires CITI (Golchay, Le Mouël, Ponge, & Stouls, 2016 ; Lèbre, 2016) et SC3 (Barrios et al., 2016), l'INSA de Lyon et l'Université Industrielle de Santander (UIS) proposent de travailler sur deux points durs garantissant ce service :

- Architecture fiable de services : (1) efficace et temps-réel, (2) distribuée à différentes échelles, (3) tolérante aux pannes et persistante
- Services citoyens intelligents : (1) autonomes, (2) pertinents, (3) sociaux, (4) participatifs - crowdsensing/sourcing.

Les sections suivantes détaillent les avancées déjà menées sur le sujet.

# Résultats

## Architecture distribuée de services

### *Découverte et infrastructure dynamique de services*

Une des questions critiques est de comment promouvoir le développement d'une plateforme urbaine de services, impliquant des acteurs et un processus réel dans une politique de citoyen intelligent. La Figure 1 montre une plateforme urbaine de services centrée sur les citoyens (Barrios et al., 2012). Bien que centralisée, cette plateforme peut être transférée vers des solutions distribuées - comme dans un Cloud Spontané de proximité (cf Figure 2) (Golchay et al., 2016).

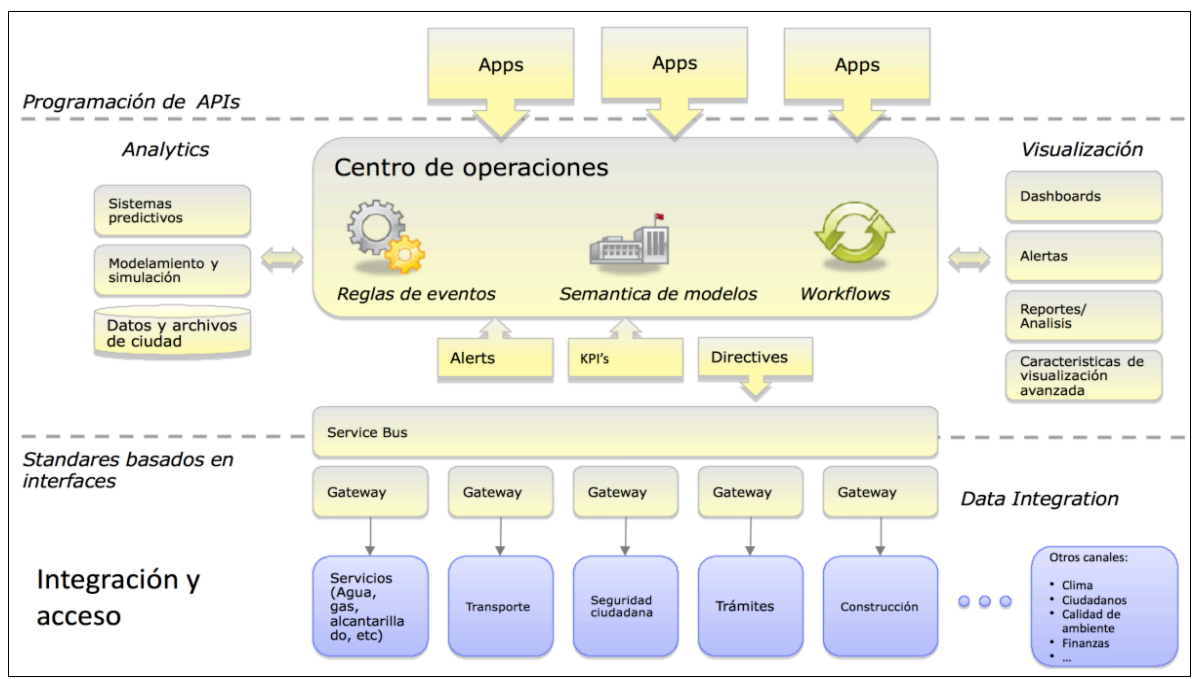

*Figure 1:* Plateforme Urbaine de Services

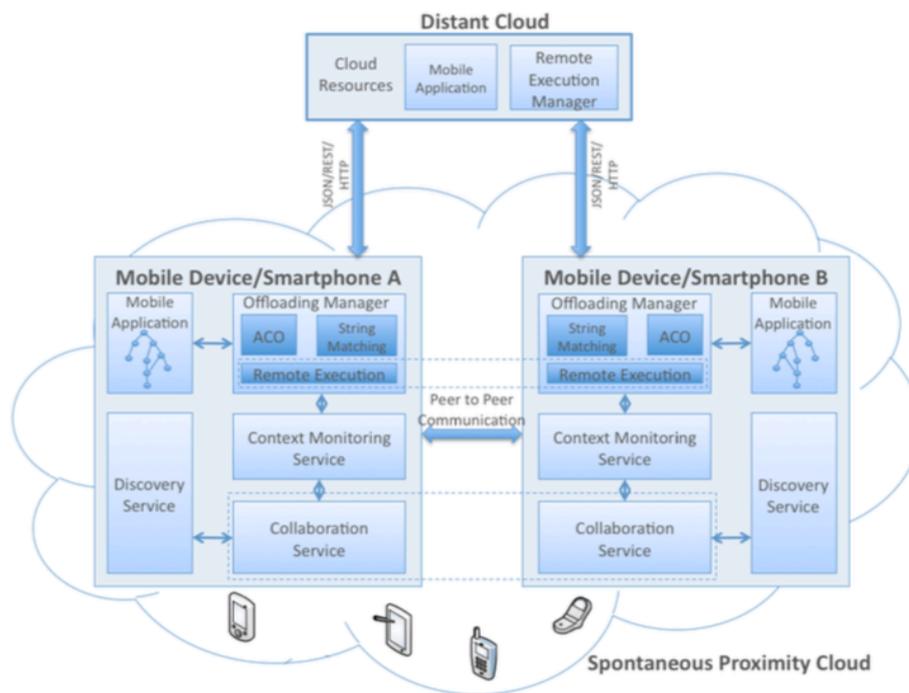

*Figure 2:* Architecture de services distribuées - Cloud Spontané de proximité

**Prise de décision collaborative**

*Algorithmes bio-inspirés du comportement des fourmis pour la gestion du trafic*

La fluidité du trafic véhiculaire est un enjeu majeur et est un exemple de service montrant bien les aspects décentralisés et collaboratifs. Nous proposons un service véhiculaire basé sur le crowdsourcing où le véhicule est assimilé à une fourmi cherchant son chemin en temps réel (cf Figure 3) (Lèbre, 2016). Au cours de leurs déplacements, les véhicules échangent leur connaissance du réseau avec les autres véhicules (cf Figure 4). Ils calculent donc leur trajet avec une information partielle du réseau.

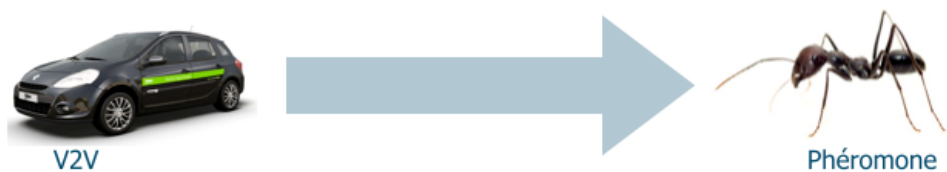

*Figure 3:* Utilisation de modèles bio-inspirés pour un service véhiculaire de trafic

Les résultats peuvent être très bons en cas de trafic normal (KPP dans la Figure 5(a)) et avec un algorithme adapté dynamiquement en cas de catastrophe - ici un tremblement de terre (PPE dans la Figure 5(b)).

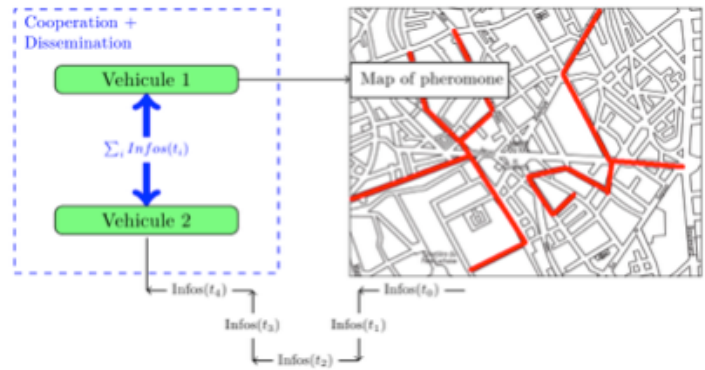

*Figure 4:* Échange de cartes contenant des données de

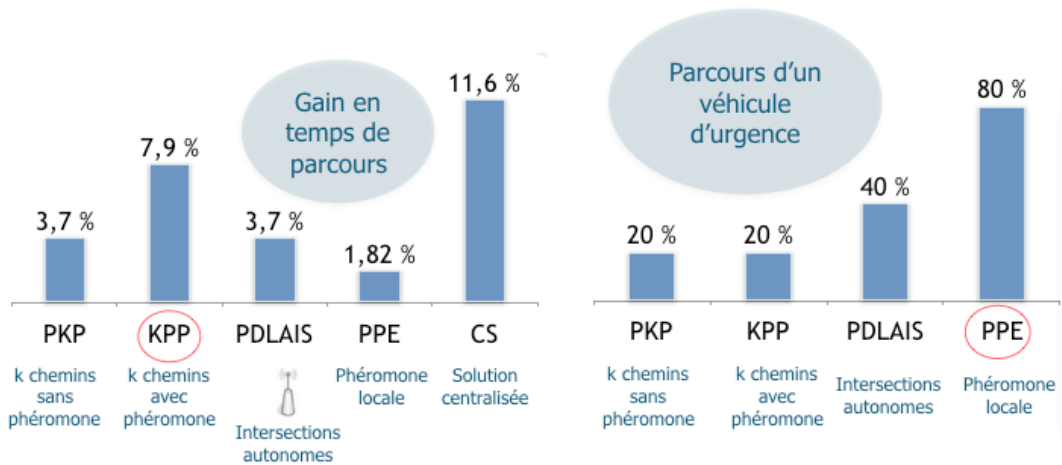

*Figure 5:* (a) Gain de temps de parcours en condition de trafic normal vs. (b) probabilité

***Assistance aux personnes et véhicules par prévision du comportement en situation d'urgence***

Une des étapes importantes du projet ALERT est de pouvoir simuler en avance la réponse attendue des services. Une des premières simulations faite a été le comportement des foules lors d'une urgence suite à un attentat par engin explosif (cf Figure 6) (Burgos, 2015).

## Méthodologie

Pour la mise en place du projet ALERT, nous proposons de continuer nos simulations sur le modèle du campus de l'UIS pour évaluer les besoins techniques et logiciels de déploiement du service. Ensuite, un passage à l'échelle dans la ville se fera en accord avec la municipalité.

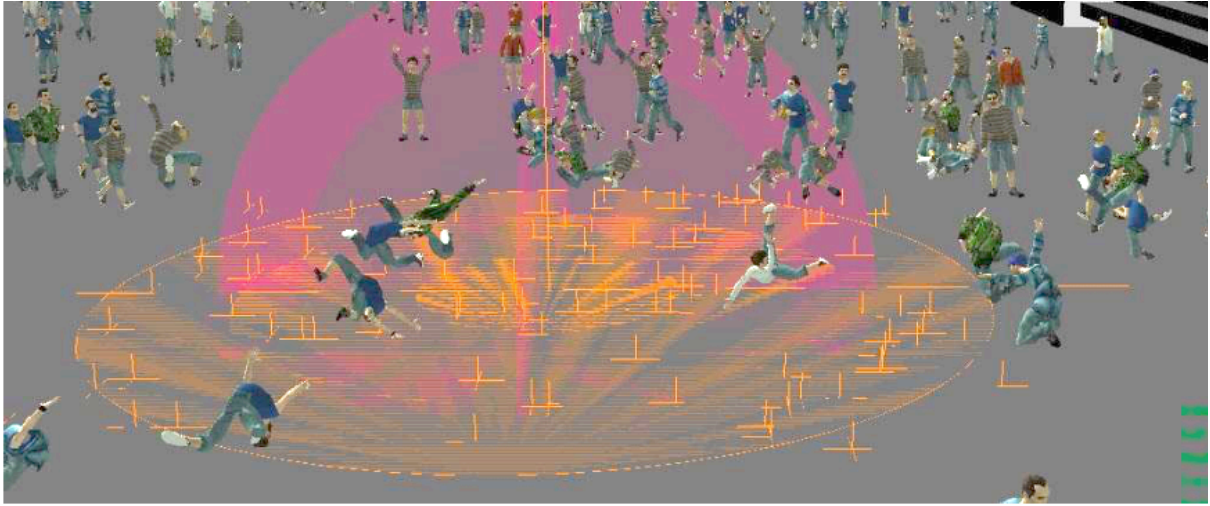

## Discussion

Une mise en place à Bucaramanga nous semble un choix intéressant, la ville étant en pleine expansion et particulièrement bien placée au niveau du déploiement numérique.

De même, le laboratoire SC3 y est installé et jouit d'une position phare en Colombie dans la gestion des données et calcul de haute performance.

## Remerciements

### Sources

Barrios, C., Pedraza, G., Hernández, J. T., Castro, H., Riveill, M., Roncancio, C., ... Denneulin, Y. (2016). Rapport d'activité de collaboration franco-colombienne cataÏ en informatique avancée pour le développement durable.

Barrios, C., Puleo, R., Cruz, J., Bedoya, D., Briceño, Y., Diaz Toro, G. J., ... Núñez de Villavicencio, L. (2012). Un Modelo de Autosostenibilidad y Servicio para Computación Avanzada en Latinoamérica inspirado en Aplicación como Servicio (AaaS). Segunda Conferencia de Directores de Tecnología Gestión de las TI en Ambientes Universitarios - TICAL 2012.

Burgos, D. (2015). Simulación y visualización de la dinámica del comportamiento de multi- tudes usando aceleradores gráficos (PhD Thesis). Universidad Industrial de Santander, Bucaramanga, Colombie.

Golchay, R., Le Mouël, F., Ponge, J., & Stouls, N. (2016, novembre). Spontaneous proximity clouds : Making mobile devices to collaborate for resource and data sharing. In Proceedings of the 12th eai international conference on collaborative computing : Networking, applications and worksharing (Collaboratecom'2016). Beijing, China. Consulté sur https://hal.inria.fr/hal-01391114 (BestPaper)

Lèbre, M.-A. (2016). De l'impact d'une décision locale et autonome sur les systèmes de transport intelligent à différentes échelles (PhD Thesis). Université de Lyon, INSA Lyon, Lyon, France.